# Transport and Magnetic properties of $Fe_xVSe_2$ (x = 0 -0.33)

## C S Yadav and A K Rastogi

School of Physical Sciences,  Jawaharlal Nehru University,
New Delhi- 110067, India.

E-mail: shekharjnu@gmail.com, akr0700@jnu.ac.in

**Abstract**

We present our results of the effect of Fe intercalation on the structural, transport and magnetic properties of 1T-$VSe_2$. Intercalation of iron, suppresses the 110K charge density wave (CDW) transition of the 1T-$VSe_2$. For the higher concentration of iron, formation of a new kind of first order transition at 160K takes place, which go on stronger for the 33% Fe intercalation.  Thermopower of the $Fe_xVSe_2$ compounds (x = 0 - 0.33), however do not show any anomaly around the transition.  The intercalation of   Fe   does not trigger any magnetism in the weak paramagnetic 1T-$VSe_2$, and Fe is the low spin state of $Fe^{3+}$.

**Introduction**

The intercalated compounds of Group Vb transition metal (V, Nb and Ta) dichalcogenides have been known to show  a variety of physical phenomenon like charge density wave (CDW), Spin density wave (SDW), superconductivity and electronic correlation effects [1-3]. The ambivalent nature viz. (i) covalency of the d electrons leads to anisotropic binding with the anions, and (ii) the remaining non bonding localized electrons show highly correlated behavior and lie around the boundary of Mott transition. These give a rich variety of the electronic phase transitions with doping, pressure, electric and magnetic field.

The intercalation of the atoms or substitutional alloying of these layered compounds affects the structural and electronic properties in different manner. Intercalation of relatively small concentration of atoms, often gives a Kondo like minimum from 10K to 30K in the resistivity in the compounds viz. 2H-$Fe_{0.05}TaS_2$, 2H-$Fe_{0.05}NbSe_2$ [3], 3R-$Nb_{1.07}S_2$ , 3R-$Ga_xNbS_2$ [4] and 1T-$V_{1.1}S_2$ and 3T-$Al_{0.05}VS_2$ [5]. The intercalation of 5% Fe atoms in 2H-$TaS_2$ increases the superconducting transition from 0.8K to 3.3K, supposedly by suppressing CDW [3], while with 25% Fe intercalation, it becomes a ferromagnetic metal below Tc ~ 160K and has large anisotropy and coercive field [6]. The substitutionally alloyed Fe in 1T-$Fe_{0.07}Ta_{0.93}S_2$ on the other hand, is diamagnetic at room temperatures. The saturating magnetization behavior at low temperature in this compound is explained as due to a mixed CDW-SDW state [7]. With the larger amount of Fe substitution in 1T-$Fe_xTa_{1-x}S_2$ (x < 0.33) compounds, a transition from the diamagnetic low spin to high spin state has been found [8].

In this present study, we have intercalated Fe in the 1T-VSe$_2$, to see the effect of charge transfer and the effect of Fe intercalation on its magnetic properties. The band structure calculations suggests, 1T-VSe$_2$ to be exchange enhanced paramagnetic material with a stoner enhancement factor S= 4.47 [9]. It is expected that the Fe intercalation may trigger ferromagnetism in this compounds like in the case of Palladium. Our studies as explained in the subsequent sections, have shown the presence of charge transfer effect in resistivity and thermopower of the Fe$_x$VSe$_2$ compound, but even up to 33% concentration of Fe intercalation, the compound remains paramagnetic.

**Sample Preparation**
Compounds were prepared by the Solid State chemical reaction inside the evacuated quartz tubes in two stages. VSe$_2$ and FeSe were first prepared separately by reaction at 700$^0$C and 550$^0$C respectively of V (99.7%) and Fe (99.98%) chips using 5% excess Se (99.999%). VSe$_2$ and FeSe were thoroughly mixed in required molar proportions, pelletized at 5 tons and then sintered at 550$^0$C for five days. The excess Se was separated out at the colder end of the tube. The low temperature of the synthesis and higher Se content was used to avoid the self- interaction of V in the van der Waal gap. Single crystal of the VSe$_2$ and Fe$_{0.33}$VSe$_2$ of dimension 3×3×0.02 mm$^3$ were also obtained by the vapor transport at 700$^0$- 650$^0$C temperature gradient. For the later part of measurements, the compounds were also prepared at higher temperature 700$^0$C, to see the effect of high temperature synthesis on the properties.

**X-Ray Diffraction and Structure Refinement**
The X-ray diffraction Pattern of the Fe$_x$VSe$_2$ (x = 0.02, 0.10, 0.15, 0.20, 0.33) prepared at 550$^0$C (low temperature phase) are shown in figure 1. As observed from the pattern, up to 20% intercalation of Fe, do not give rise to any new peaks, except for the expected changes in the relative intensity. The slight broadening of lines and the additional weak peaks in the 33% Fe intercalated compound may be related to structural distortion due to the short range ordering of the Fe atoms. The lattice parameters of these compounds were calculated using Crysfire software and are listed in the table-1 [10-11].

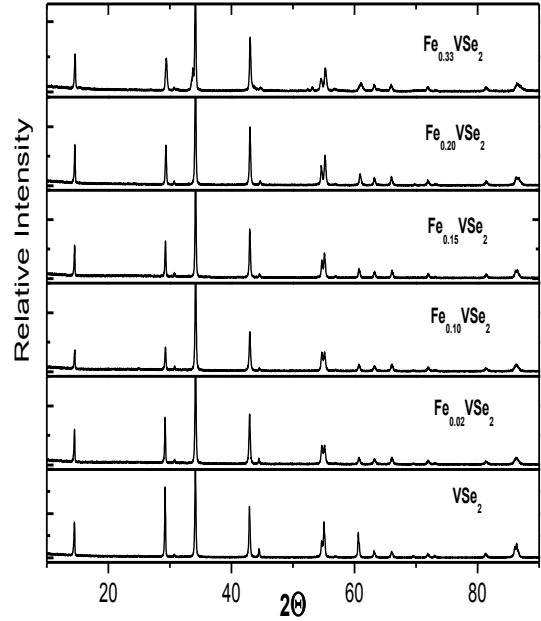

Figure 1. XRD pattern of the low temperature phase of Fe$_x$VSe$_2$ (x = 0, 0.02, 0.10, 0.15, 0.20, 0.33).

**Table -1**

| Fe$_x$VSe$_2$ x | a(Å) | b (Å) | c (Å) | V (Å$^3$) | Comment |
|---|---|---|---|---|---|
| **0.00** | 3.356 | 3.356 | 6.104 | 68.75 | Hexagonal |
| **0.02** | 3.356 | 3.356 | 6.100 | 59.54 | Hexagonal |
| **0.10** | 3.351 | 5.814 √3×3.567 | 6.097 | 118.77 Z = 2 | Orthorhom |
| **0.15** | 3.357 | 5.808 √3×3.353 | 6.097 | 118.86 | Orthorhom |
| **0.20 (LT)** | 3.359 | 5.764 √3×3.328 | 6.081 | 117.75 | Monoclinic |
| **0.20 (HT)** | 3.351 | 3.351 | 6.090 | 59.234 | Hexagonal |
| **0.33 (LT)** | 3.527 | 5.819 √3×3.360 | 2×6.074 | 249.6 γ = 91.61 | Monoclinic |
| **0.33 (HT)** | 3.365 | 3.365 | 6.056 | 59.36 | Hexagonal |

As shown in table-1, for the 2% Fe intercalated VSe$_2$, the lattice parameters are c = 6.100 Å and a = 3.356 Å, remain unchanged from that of 1T-VSe$_2$ (c = 6.104 Å and a =3.356 Å). For the other concentrations, structures start deviating from the hexagonal, to orthorhombic and monoclinic one. The decreasing 'c' parameter and increasing 'a' parameter leads to the consistent reduction in the c/a value.

For low temperature phase 33% Fe intercalated compound (LT), the ordering of the Fe lead to the superstructure formation with lattice parameters a = 3.527Å, b = 5.819Å (√3×3.360) and c = 12.14Å (2×6.07). The high temperature synthesis, stabilizes the hexagonal structure, and even for the 33% Fe intercalated phase, does not shown any distortion unlike its low temperature counterpart prepared at $550^0C$.

**Transport Properties**

*Electrical Resistivity*
We measured the electrical resistivity of the $Fe_xVSe_2$ compounds from 2K to 300K. The resistivity of the compounds, after normalizing at room temperature is shown in figure 2. It is seen form the figure that the temperature dependence of the resistivity varies from the metallic like to semimetal/semiconductor like nature. The 110K CDW transition, observed in $VSe_2$, get diminished for more than 5% intercalation. The CDW onset temperature $T_o$ (where the resistivity deviates from the linear dependence), reduces for 5% or less Fe concentration, But for more than 10% Fe concentration, a new kind of first order transition appears at 160K, and show a hysteresis while cooling and warming. This hysteresis becomes broader for the higher concentration of Fe.

In the figure 3, we have plotted the resistivity of the $Fe_xVSe_2$ polycrystalline pellets as well as the $Fe_xVSe_2$ crystal flakes, in the upper and lower panel of the figure respectively. In the upper panel, the resistivity of the two high temperature phases of $Fe_{0.20}VSe_2$ and $Fe_{0.33}VSe_2$ which were prepared at $700^0C$ are also plotted. We did not see any hysteresis behavior for $Fe_{0.20}VSe_2$ (HT), and also in case of the $Fe_{0.33}VSe_2$ (HT), the 160K transition gets considerably reduced and the hysteresis get confined to a narrow temperature range, in comparison to the low temperature phase of $Fe_{0.33}VSe_2$, which was prepared at $550^0C$. Both these high temperature phases show absence of distortion of the parent hexagonal structure. We can conclude from this behavior, along with our structural studies of both the Low temperature

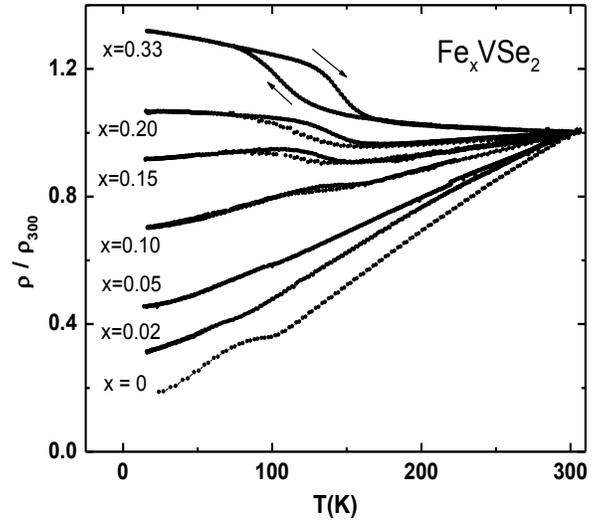

Figure 2 Temperature dependence of the Normalized resistivity of polycrystalline $Fe_xVSe_2$ (x = 0, 0.02, 0.05, 0.10, 0.15, 0.20, 0.33) compounds.

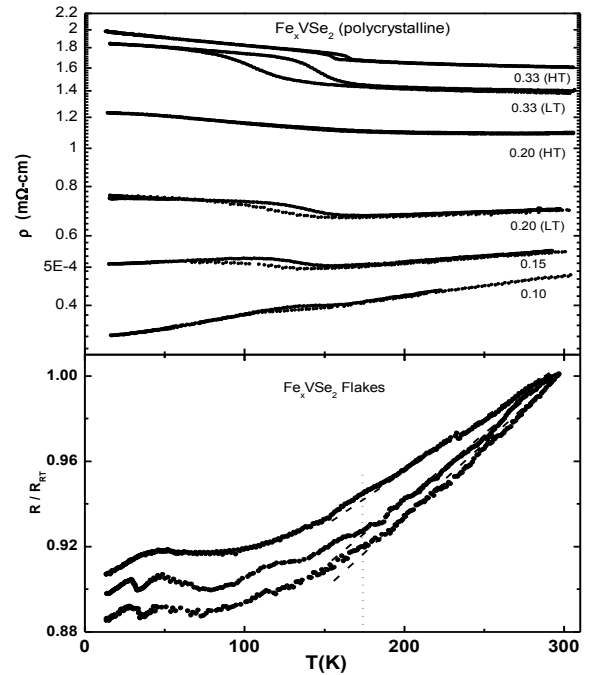

Figure 3. Electrical resistivity of polycrystalline $Fe_xVSe_2$ (x = 0, 0.10, 0.15, 0.20, 0.33) (upper

panel) and $Fe_xVSe_2$ flakes (lower panel), showing the anomaly around 160K transition.

(LT) and high temperature (HT) phases of $Fe_{0.20}VSe_2$ and $Fe_{0.33}VSe_2$, that the high temperature synthesis of the compounds reduces the distortion of structure, which in turns suppresses the 160K transitional anomaly in conductivity properties.

The normalized resistivity of the $Fe_xVSe_2$ single crystals is shown in the lower panel of the figure 3. These flakes were obtained by the vapor transportation of $Fe_{0.33}VSe_2$ composition. However, we could not ascertain the actual concentration of Fe in the flakes. All the three measurements shown in the figure were done on the crystal flakes obtained from the same batch. All the crystal flakes show metal like conductivity, unlike the case of polycrystalline pellets, where resistivity rises on lowering the temperature. This shows a highly anisotropic nature of the compounds, as in case of crystal flakes, we measure the conductivity along the a-b plane. Since the parent compound $1T-VSe_2$ is a layered material, with a weak van der Waal bonding between the layers. Hence c-axis properties (like conductivity) may be quite different from those along the a-b plane. Similar features are also observed for other layer materials, like graphite or high Tc superconductors.

Though in the flakes of $Fe_xVSe_2$, a metal like conductivity is found, the presence of 160K transition can also be noticed. The resistivity of the compounds starts deviating from their linear temperature dependence after 160K on cooling. This is shown with the help of some dotted lines in the figure 3.

On further cooling, the resistivity of the crystal flakes shows a sharp downward trend after 50K. At present, we have no explanation for this type of behavior.

*Thermoelectric Power*

Thermoelectric Power (TEP) of the polycrystalline $Fe_xVSe_2$ compounds is shown in the figure 4. For all the compositions, thermopower varies from $2\mu V/K$ at 15K to a saturated value of $13\mu V/K$, near room temperatures. In the lower panel of the figure 4, we have shown the TEP of 2% and 5% Fe intercalated compounds, which were prepared at two different temperatures $700^0C$ and $550^0C$, marked as HT and LT respectively, in the figure. These compositions showed a shift in the CDW transition to lower temperatures as discussed in previous section.

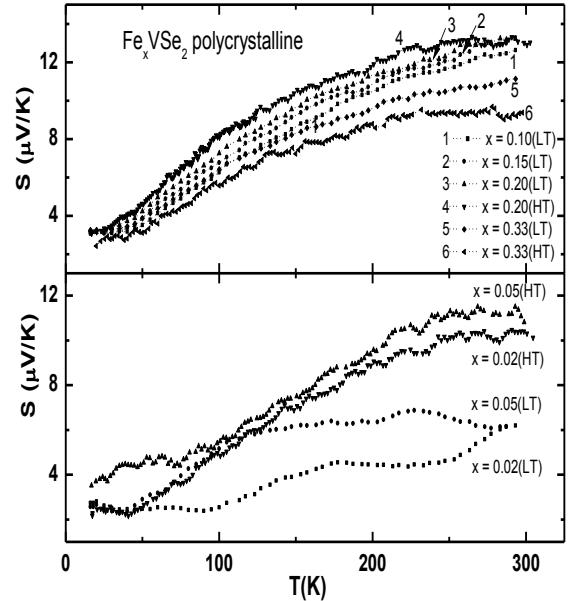

Figure 4. Temperature dependence of the thermopower of polycrystalline $Fe_xVSe_2$; x = 0.02, 0.05 (lower panel) and x = 0.10, 0.15, 0.20, 0.33. HT and LT written in the parentheses denote the high temperature and low temperature phase of the compounds respectively.

As seen from the figure, TEP of the high temperature phases has higher value. The high temperature phases of 2% and 5% are very similar to their higher concentration counterpart, except for a sharp change in slope at the temperatures around 70K.

For the high concentration phases shown in the upper panel of figure 4, TEP vary in a similar manner for all compositions, irrespective of their preparation condition also. Ironically, the TEP for $Fe_{0.33}VSe_2$, lies to the lowest values, may be due to the possible ordering of Fe atoms, because of 1/3 sites being filled. This ordering can be instrumental in distorting the topology of Fermi surface, and can show the changes in the thermopower. For rest of other compositions (x = 0.10, 0.15, and 0.20), thermopower value consistently increases with the concentration, and thus show the effect of charge transfer from intercalate Fe to the host V atom.

Thermoelectric power of the metal, where the system of electrons interact with a random distribution of scattering centers, which are in thermal equilibrium at the local temperature T, is given by

$$S = \frac{\pi^2}{3} \cdot \frac{k^2}{e} T \cdot \left(\frac{\partial \ln \sigma}{\partial \varepsilon}\right)_\eta$$

Where, $k$ = Boltzmann constant, e = charge of electron, $\sigma$ = electrical conductivity of electrons of energy $\varepsilon$, and $\eta$ is the chemical potential of electrons the higher Fe concentration (x = 0.10 - 0.33) samples. Whereas for the low concentration (x = 0.02-0.05) compounds, the low temperature phase of 2% and 5% Fe intercalated compounds show similar behavior and S/T remain nearly constant till 40K. We have also shown the TEP of $Fe_xVSe_2$ crystal flake and the $VSe_2$ crystal flakes in the figure 6. The TEP of $Fe_xVSe_2$ crystal neither show 110K CDW transition as in the $VSe_2$ flake, nor does it show any signature of 160K. Instead it shows a pure metallic behavior down till 30K.

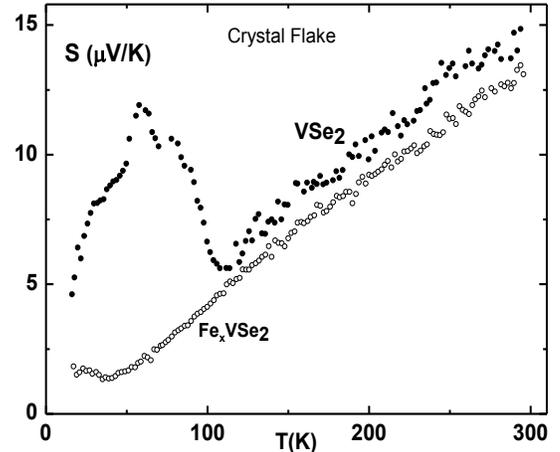

Figure 6. Thermopower of the 1T-$VSe_2$ and $Fe_xVSe_2$ crystal flakes.

## Magnetic Properties

We have measured field dependence of the magnetization of $Fe_xVSe_2$ (HT); x = 0.02, 0.05, 0.20 and 0.33; powder samples up to 14 Tesla field at 2K temperature. Our results are shown in the figure 7. The magnetization of the compounds do not show any sign of saturation, even at 14 Tesla field. This type of behavior can be possible, if some of Fe give rise to the local moments and also add to the band magnetism.

In the figure 8, we have shown the temperature dependence of the susceptibility measured at 1 Tesla magnetic field. The data are fitted to the Curie Weiss formula

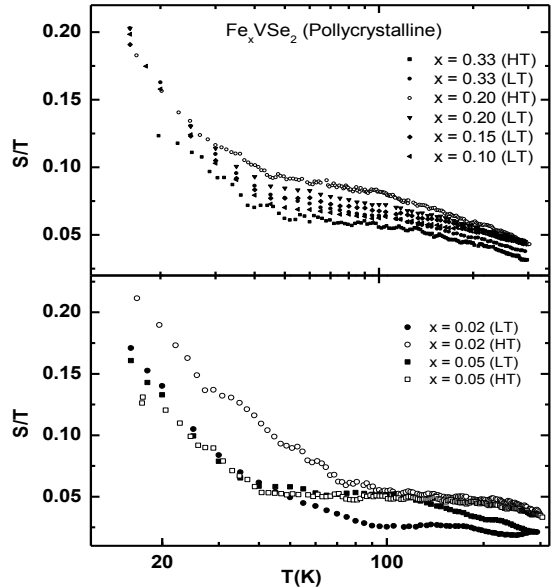

Figure 5. Temperature dependence of the thermopower of polycrystalline $Fe_xVSe_2$ plotted as S/T. x = 0.02, 0.05 (lower panel) and x = 0.10, 0.15, 0.20, 0.33.

We have plotted S/T as a function of temperature, in figure 5, to show the behavior of the derivative of the logarithmic conductivity, with respect to the energy, at the Fermi surface. As seen from the upper panel in figure 5, S/T rises linearly, on cooling and changes its slope in the region 80K-170K, for

$$\chi = \chi_0 + \frac{C}{T+\theta}$$

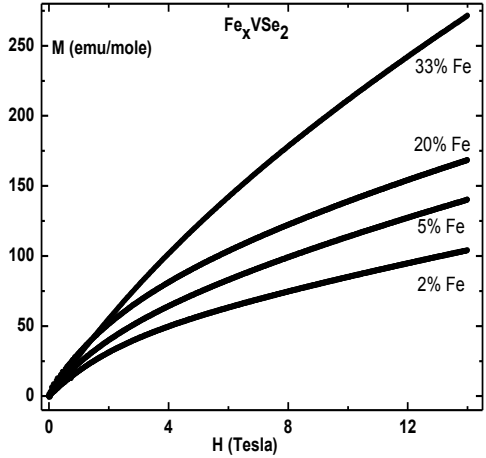

Figure 7. Magnetic field dependence of magnetization of $Fe_xVSe_2$ at 2K, showing non-saturating behavior up to 14 Tesla field.

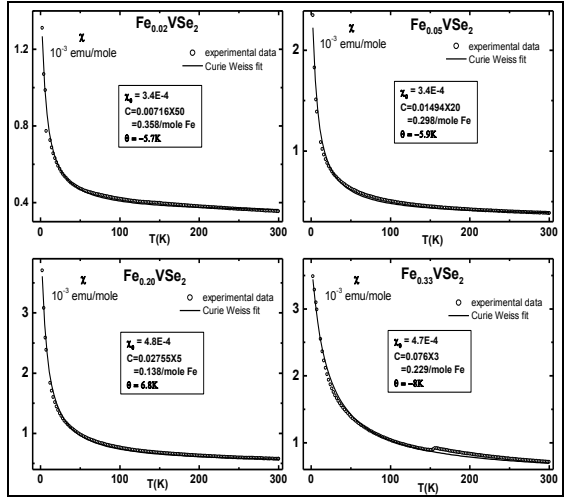

Figure 8. The temperature dependence of the susceptibility of the $Fe_xVSe_2$, measured at 1Tesla magnetic field. Curie Weiss fit of the data is also shown in the figure.

The values of different parameters; $\chi_0$ (the temperature independent contributions to susceptibility), C (Curie constant) and θ (antiferromagnetic Neel temperature), obtained from the fitting are also shown in the figure 8. As shown in the table 2, the value of the effective magnetic moment per Fe atom decreases from $1.69\mu_B$ to $1.05\mu_B$ for Fe concentration 2% to 20%. But for the 33% Fe intercalated compound, it is found to be $1.35\mu_B$. We also observe a small jump like anomaly in the susceptibility of $Fe_{0.33}VSe_2$ at 160K transition. This seems to be related to the anomalies in the resistivity of the compound at the same temperatures. The magnetic behavior of $Fe_{0.33}VSe_2$ has been discussed in detail in the separate study [12].

Table- 2

| Compound | $\chi_0$ $10^{-4}$ emu/mole-K | $C_{per Fe}$ emu/mole-K | $\mu_{eff}$ / Fe $\mu_B$ | θ K |
|---|---|---|---|---|
| $Fe_{0.02}VSe_2$ | 3.4 | 0.36 | 1.69 | -5.7 |
| $Fe_{0.05}VSe_2$ | 3.4 | 0.30 | 1.55 | -5.9 |
| $Fe_{0.20}VSe_2$ | 4.8 | 0.14 | 1.05 | -6.8 |
| $Fe_{0.33}VSe_2$ | 4.7 | 0.23 | 1.35 | -8.0 |

Low value of the magnetic moment indicates that the Fe atoms are in the low spin state. Similar results were also obtained by DiSalvo et.al. for the Fe substituted $VSe_2$ (1T- $V_{1-x}Fe_xSe_2$), where magnetic moment per Fe atom was $0.7\mu_B$ and $0.6\mu_B$ for x = ½ and 1/3 respectively [13]. In our intercalated Fe compounds, the magnetic moments are much higher but still in the low spin state regime.

**Conclusion**

We observed the effect of Fe intercalation on the properties of the $1T-VSe_2$ compound. The intercalation of Fe suppresses the 110K CDW transition, and a new first order transition appears at 160K, which gets stronger for higher Fe intercalation especially for the phases prepared at low temperatures. The resistivity of $Fe_xVSe_2$ crystal flakes shows metallic behavior, contrary to that of the polycrystalline samples, which shows rise in resistivity on cooling. The positive temperature coefficient of the resistance of the flake contrast with the negative temperature coefficient of corresponding polycrystalline samples. In polycrystalline sample the resistivity is due to combination of in-plane and c-axis resistivity of the anisotropic structures. This shows that the anisotropy of the transport increases on cooling, i.e. there is increasing localization of wavefunction of conduction electron in the direction perpendicular to the layers as temperature decreases and also as concentration of the intercalated atom increases. The Thermopower of the compounds do not show any anomaly around the transition and continuously rises on warming before saturating near room temperature, and indicates the transfer of electron from Fe to Vanadium band. Like the resistivity of the $Fe_xVSe_2$ flake, thermopower of the crystal flake also show metallic behavior and shows highly anisotropic nature of the compounds.

$Fe_xVSe_2$, do not show any saturation even for 14Tesla magnetic and remain a weak paramagnetic, and cannot be understood in the Heisenberg model of localized moment. The other picture of itinerant band magnetism seems more appropriate for this system. As mentioned earlier that we started with the presumption that the presence of Fe atom may trigger the magnetism in the narrow d- band of correlated $1T-VSe_2$ compound, But it did not happened, and for even 50% substitution or 33% intercalation of Fe, does not produce ferromagnetism in the compound [13]. Like the magnetic behavior of the Fe substituted $VSe_2$, our magnetic study also indicate the Fe to be in the low spin state for the Fe intercalated $VSe_2$ [13]. So the possibility of induced ferromagnetism in the $VSe_2$, is negated at least for the up to 33% Fe concentration. The effective magnetic moment per Fe atom was found as $1.69\mu_B$, $1.55\mu_B$, $1.05\mu_B$, and $1.35\mu_B$ for the $Fe_xVSe_2$; x = 0.02, 0.05, 0.20 and 0.33 respectively. The crystal field effect calculation on these compounds can give the detailed picture of the magnetic moment variation.

**Acknowledgements**
CSY acknowledges Council of Scientific and Industrial Research (CSIR) India for the Senior Research fellowship.

**References**

[1] R H Friend, A R Beal and A D Yoffe, 1977, Phil Mag, **35-5**, 1269
[2] S S P Parkin and R H Friend, 1980, Phil Mag. 41-1 65, *II. Transport properties,* Phil Mag. **41-1** 95, *III. Optical properties*, Phil Mag. **42-5** 627
[3] D A Whitney, R M Fleming and R V Coleman, 1977, Phys. Rev. B**, 15**, 3405
[4] A. Niazi and A K Rastogi, 2001, J. Phys.: Condens. Matter **13**, 6787
[5] P Poddar and A K Rastogi, 2002, J. Phys.: Condens. Matter. **14**, 2689
[6] E Morosan, H W Zandbergen, Lu Li, M Lee, J G Checkelsky, M Heinrich, T Siegrist, N P Ong and R J Cava, 2007, Phys Rev B **75**, 104401
[7] S Dai, C Yu, D Li, Z Shen, S Fang and J Jin, 1995, Phys. Rev. B **52**, 1578
[8] M. Eibsuitz and F. J. Disalvo, 1976, Phys Rev Lett. **36**, 104
[9] H W Myron, 1980, Physica **99B**, 243
[10] R Shirley, *The crysfire 2002 system for Automatic Powder Indexing: User's Manual*, The Lattice press, 41, Guildford Park Avenue, Guildford surrey, GU27NC, England
[11] D Tauplin, 1979, J. Appl. Cryst. **6**, 380
[12] C S Yadav and A K Rastogi, submitted to Journal of Physics: Condens Matter
[13] F J DiSalvo and J V Waszczak, *1976,* Journal de Physique, **37**, C4-157